\pdfoutput = 1
\documentclass[margin=0.5in]{article}
\usepackage[left=3cm, right=3cm, top=3cm]{geometry}
\usepackage{graphicx}
\usepackage{float}
\usepackage[mathscr]{euscript}
\newcommand{\KK}{${\mathscr{KK}}$}
\newcommand{\AFB}{$A_{\rm FB}$}
\begin{document}
\begin{titlepage}

\begin{flushright}
14th International Symposium on Radiative Corrections (RADCOR2019)\\
9-13 September 2019, Palais des Papes, Avignon, France
\end{flushright}

% ---------------- Title  ----------------
\vspace{5mm}
\begin{center}
{\Large\bf ISR and IFI in Precision AFB Studies with KKMC-hh${}^{\star}$}
\end{center}

% ---------------- Authors ----------------
\vskip 5mm
\begin{center}
{\large S.A.\ Yost${}^{a}$, S.\ Jadach${}^b$, B.F.L. Ward and Z.\ W\c{a}s}

\vskip 2mm
{\em ${}^a$Institute of Nuclear Physics, Polish Academy of Sciences,\\
  ul.\ Radzikowskiego 152, 31-342 Krak\'ow, Poland}\\
\vspace{1mm}
{\em ${}^b$The Citadel, Charleston, SC, USA\\ }
\end{center}

\vspace{2mm}
\begin{abstract}
{\KK}MC-hh is a hadronic event generator for Z boson production and 
decays, which includes exponentiated multi-photon radiation and first-order 
electroweak corrections. We have used {\KK}MC-hh to investigate the role of 
initial sate radiation (ISR) and initial-final interference (IFI) in precision 
electroweak analyses at the LHC. We compare the effect of this radiation on 
angular distributions and forward-backward asymmetry, which are particularly 
important for the measurement of the weak mixing angle.  We discuss the 
relation of the ISR implementation in {\KK}MC-hh to ISR from parton distribution
functions with QED corrections.
\end{abstract}

\vspace{50mm}
\footnoterule
\noindent
{\footnotesize
${}^{\star}$This is an extended version of the proceedings article which adds comparisons 
of results with and without a parton shower. These comparisons were included in the slides
but omitted in the published version of the proceedings article. Some comments on DIZET and
TAUOLA updates are also added here.\\
${}^{a}$This presentation was supported by a grant
from The Citadel Foundation. Computational resources were provided by the
Institute of Nuclear Physics, IFJ-PAN, Krakow, Poland.\\
${}^{b}$This work was supported in part by Polish National Science Center 
grant 2016/23/B/ST2/03927.
}

\end{titlepage}

\section{Introduction}
Angular distributions for $pp\rightarrow Z/\gamma^*\rightarrow$ leptons are 
important for a precision measurement of the weak mixing angle at the LHC.  
The inputs for calculating the weak mixing angle can come from measurements 
of the forward-backward asymmetry \AFB\ or the angular coefficient 
$A_4=4\langle \cos\theta\rangle$. In either case, the relevant angle is 
taken to be the Collins-Soper (CS) angle in the rest frame of the 
final state lepton pair.\cite{CollinsSoper} 

The angular distribution is sensitive to radiative corrections. In the presence
of final state radiation (FSR) from the leptons, the photon momenta can be 
subtracted to find the CM momentum of the Z boson. Initial state radiation 
(ISR) complicates this because it cannot be unambiguously distinguished from 
FSR. ISR also interferes with FSR at the quantum level, and this initial-final
interference (IFI) creates an ambiguity in the Z boson rest frame that cannot
be resolved, even in principle.  These radiative effects are presently 
under investigation using a variety of programs in addition to {\KK}MC-hh, 
including POWHEG-EW\cite{PowhegEW} and MC-SANC\cite{MCsanc1,MCsanc2}.

We present studies of radiative corrections to angular distributions using 
{\KK}MC-hh\cite{kkmchh}, a hadronic event generator based on 
CEEX\cite{Jadach:2000ir}, an amplitude-level soft photon exponentiation scheme 
originally developed for electron-positron collisions in the LEP era, which 
implemented for $e^+e^-$ scattering in the {\KK}MC 
generator\cite{Jadach:1999vf} and extended to 
quark initial states in {\KK}MC 4.22\cite{Jadach:2013aha}.
CEEX is similar to YFS soft photon exponentiation\cite{yfs}, but implemented
at the amplitude level rather than the cross section level, which facilitates
the exponentiation of interference effects, in particular IFI. An extensive
review and explanation of the implementation of IFI in the CEEX framework
can be found in Ref. \cite{JadachYost2019}. 

\KK{MC}-hh events can be exported in an LHE-compatible~\cite{lhe-format} 
event record and processed by an external QCD shower generator, or they
can be showered internally via HERWIG 6.5\cite{HERWIG}.
This assumes an approximation in which QCD and QED effects factorize, which
is true at leading log and should be a good approximation at 
${\cal O}(\alpha_s\alpha)$.\cite{den-ditt1211.5078,dittmr}
Mostly unshowered results will be presented here, since 
the number of events needed to see the effect of radiative corrections on 
\AFB\ or $A_4$ is on the order of $10^9$ or more, requiring substantial 
computer resources, especially in the presence of the shower. A smaller 
sample of showered events was generated, but only with ISR included. Those
results will be shown only in the context of some IFI comparisons.

{\KK}MC-hh includes an {\em ab initio} calculation of QED radiation 
including quark masses, so that the results are finite in the collinear limit.
This differs from other programs capable of addressing ISR effects in
hadron scattering, such as POWHEG-EW, MC-SANC, Horace \cite{Horace1,Horace2},
and ZGRAD2 \cite{ZGRAD2}, which factorize collinear QED radiation 
with the assumption that its effect is included in the parton distribution 
functions (PDFs). Factorizing the collinear QED has the advantage
of avoiding the issue of quark masses, but setting a high factorization scale
could limit the ability to address non-collinear ISR. Also, such factorization
is not readily combined with CEEX soft photon exponentiation in {\KK}MC-hh. 

Including quark masses in the calculation raises the question of what value 
should be assigned to them. The first parton distributions to include QED 
corrections was the MRST QED PDF set\cite{MRST}, which assumed current quark
masses. This is consistent with the expectation that for deep inelastic
scattering, the colliding quarks couple perturbatively to the spectator quarks,
so that the recoil when a photon is emitted should be governed by the current
quark mass, not the constituent mass. However, some controversy remains on this
issue, which was addressed in a study\cite{KKMChh2019} applying 
{\KK}MC-hh to LHC phenomenology relevant to the $W$ mass measurement by varying
the quark masses. The mass dependence is logarithmic, so varying the light quark
masses by a factor of 10 only changes the ISR contribution by about 10\%. Since
ISR typically contributes at the order of around 0.1\% for most distributions,
the mass dependence is usually insignificant. 

\section{The Effect of Initial-State QED Radiation on Angular Distributions}

In this section, we focus on CS angle
distributions, particularly \AFB\ and $A_4$, and compare the effect of 
including QED corrections in the PDFs to the effect of adding ISR 
via {\KK}MC-hh. All results are from {\KK}MC-hh runs without a QCD shower,
producing $5.7\times 10^9$ muon events at 8 TeV. Since {\KK}MC-hh includes
collinear ISR, it must be used with pure-QCD parton distributions. These runs
use NNPDF3.1\cite{nnpdf3.1} $(\alpha_s(M_Z) = 0.21018)$. 
For comparison, we also show results for {\KK}MC-hh with ISR off, 
but with a NNPDF3.1luxQED\cite{nnpdfluxqed} parton distribution functions,
which include LuxQED photon ISR\cite{LUXqed}.

NLO electroweak corrections are added using DIZET 6.21\cite{zfitter6:1999}, 
which uses an input scheme with parameters $G_\mu$, $\alpha(0)$, and $M_Z$. 
The quark masses in DIZET are selected internally based on the vacuum 
polarization option, for which the default fit is used. Photonic radiative 
corrections are calculated using $\alpha(0)$ and PDG values\cite{PDG2018}
for the quark current masses. Otherwise, all parameters are consistent with 
the LHC electroweak benchmark study, Ref.\ \cite{benchmark}.

All results include dilepton mass cut 60 GeV $< M_{ll} < $ 116 GeV, including
those labeled ``uncut.'' The ``cut'' results include an additional constraint
$p_{\rm T} > $ 25 GeV on the transverse momentum of each muon, and $|\eta|< 2.5$
on the pseudorapidity of each muon. The forward-backward asymmetry \AFB\ is 
calculated from the cut events, while $A_4$ is calculated using uncut events. 
Final state radiation is included in all cases. Initial-final interference 
(IFI) is not included. IFI effects are discussed separately in the next section. 
In Table 1, the column labeled ``No ISR'' have ISR turned off in {\KK}MC-hh and
use a non-QED NNPDF3.1 set. The LuxQED column has ISR turned off in 
{\KK}MC-hh and uses the NNPDF3.1luxQED. The ``{\KK}MC-hh ISR'' column has ISR
turned on in {\KK}MC-hh and uses a non-QED NNPDF3.1 set. Differences are shown
comparing ISR on and off both ways, using LuxQED or {\KK}MC-hh. 
In the case of the cross-section, the differences are shown as percentages, 
while for the asymmetries, the straight differences are shown. 

%\hspace{-1.5cm}
\vbox{
%\noindent\makebox[\textwidth]{%
\begin{center}
\resizebox{\textwidth}{!}{%
\begin{tabular}{|l|c|c|c|c|c|}
\hline
& No ISR	& LuxQED ISR	& LuxQED$ - $no ISR& {\KK}MC-hh ISR & ISR$ - $no ISR\\% & ISR$ - $LuxQED\\
\hline
Uncut $\sigma$ 	&
939.858(7) pb	& 944.038(7) pb	& 0.445(1)\%	& 944.99(2) pb	& 0.546(2)\%	\\%& 0.101(2)\%\\
Cut $\sigma$	&
439.103(7) pb	& 440.926(7) pb & 0.415(1)\%	& 442.36(1) pb  & 0.742(3)\%	\\%& 0.326(3)\%\\
\AFB		&
0.01125(3)	& 0.01145(2)	& $(1.9\pm0.3)\times 10^{-4}$ 	& 0.1129(2)	&$(3.9\pm2.8)\times 10^{-5}$	
\\%&$(-1.5\pm0.3)\times 10^{-4}$\\
$A_4$		&
0.06102(4)	& 0.06131(3)	& $(2.9\pm0.5)\times 10^{-4}$	&0.06057(3)	&$(-4.4\pm0.5)\times 10^{-4}$
\\%&$(-7.4\pm0.5)\times 10^{-4}$\\
\hline
\end{tabular}
}
\\[1em]
{{\bf Table 1.} Effect of ISR added via LuxQED or {\KK}MC-hh}
\end{center}
}

Both LuxQED and {\KK}MC-hh show that ISR shifts the cut and uncut cross-section
by about half a percent, with differences on the order of a per-mil. LuxQED also shows a shift in \AFB\ and $A_4$ on the order of a few per-mil, but 
the ISR effect in {\KK}MC-hh is much smaller for \AFB, and has the opposite
sign for $A_4$.

Figures 1 and 2 compare Collins-Soper angular distributions 
$\cos(\theta_{\rm CS})$ in three cases: ``FSR only'' has no ISR and a non-
QED PDF set, ``FSR + ISR'' includes {\KK}MC-hh ISR with a non-QED PDF set, and
``FSR + LuxQED'' uses a LuxQED PDF set with no ISR from {\KK}MC-hh. Fig. 1
does not include the additional lepton cuts, and is the distribution relevant to
$A_4$, while Fig. 2 includes the lepton cuts, and is relevant to \AFB. 

\begin{figure}[h!]
\begin{center}
\setlength{\unitlength}{1in}
\begin{picture}(6.5,3.0)
\put(0.0,0.5){\includegraphics[width=3.2in,height=2.6in]{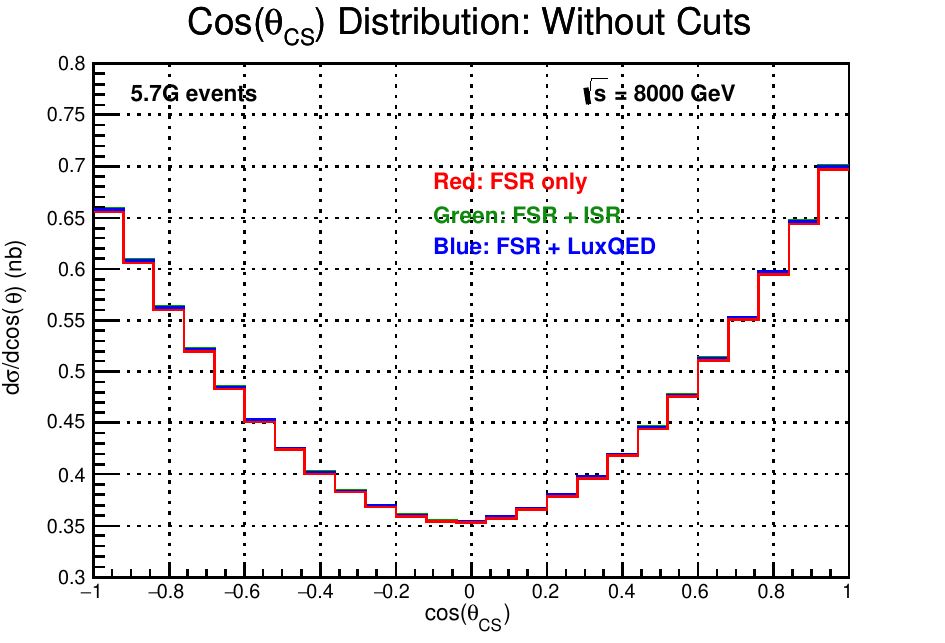}}
\put(3.0,0.5){\includegraphics[width=3.2in,height=2.6in]{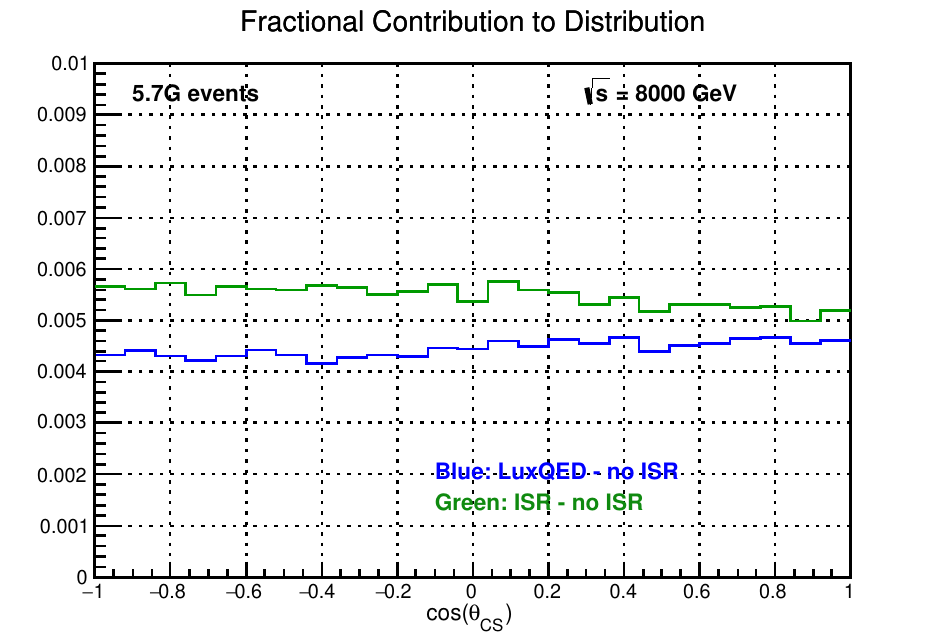}}
\end{picture}
\vspace{-0.75in}
\caption{ISR contributions to $\cos(\theta_{\rm CS})$ distributions, 
without lepton cuts.}
\end{center}
\end{figure}

\begin{figure}[hb!]
\begin{center}
\setlength{\unitlength}{1in}
\begin{picture}(6.5,3.0)
\put(0,0.5){\includegraphics[width=3.2in,height=2.6in]{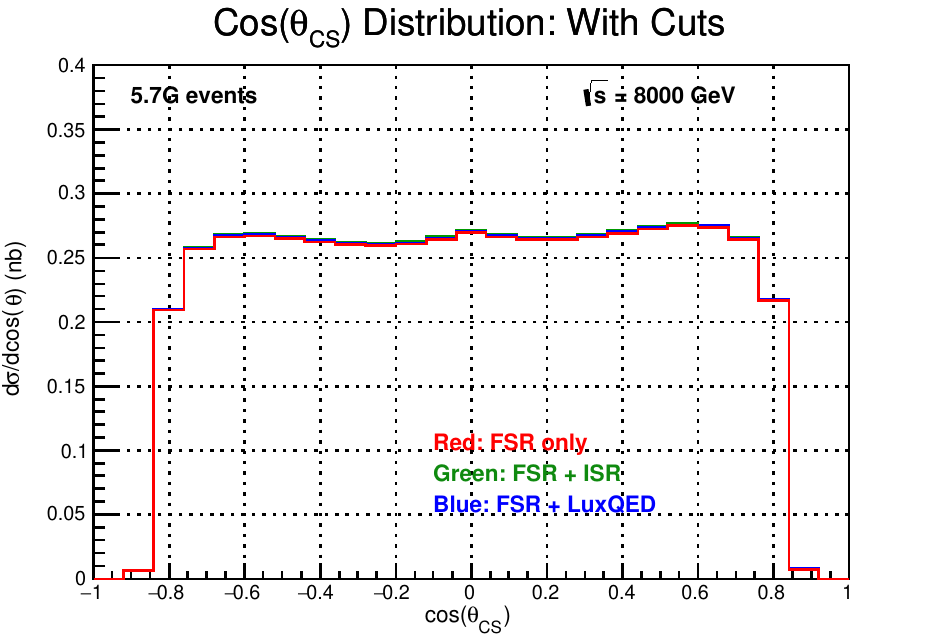}}
\put(3.0,0.5){\includegraphics[width=3.2in,height=2.6in]{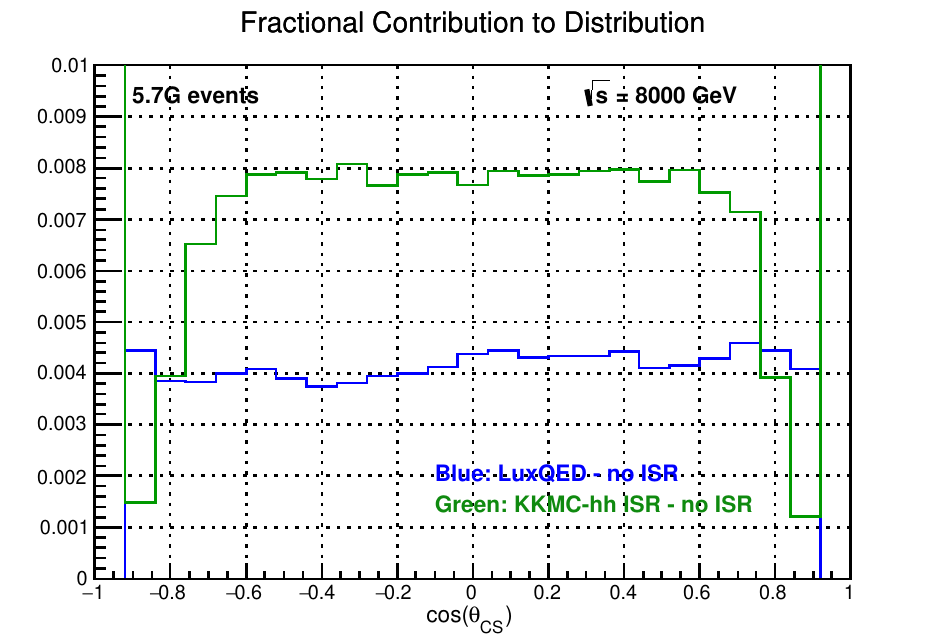}}
\end{picture}
\vspace{-0.75in}
\caption{ISR contributions to $\cos(\theta_{\rm CS})$ distributions, with lepton cuts.}
\end{center}
\end{figure}

Figures 3 and 4 show the effect of ISR on \AFB\ as a 
function of the dilepton mass and rapidity, respectively. 
In Fig. 3, the ISR contribution to \AFB\ is less than $10^{-3}$ for the entire 
range of $M_{ll}$, and in the vicinity of $M_Z\approx 91$ GeV, it is less than 
$3\times 10^{-4}$, for both LuxQED and {\KK}MC-hh. In Fig. 4, the ISR effect
from {\KK}MC-hh is below $10^{-4}$ in all bins, and consistent with zero in 
the central bin. However, LuxQED would give a larger ISR contribution
for $Y_{ll}<2$. 

\begin{figure}[H]
\begin{center}
\setlength{\unitlength}{1in}
\begin{picture}(6.5,3.0)
\put(0,0.5){\includegraphics[width=3.2in,height=2.6in]{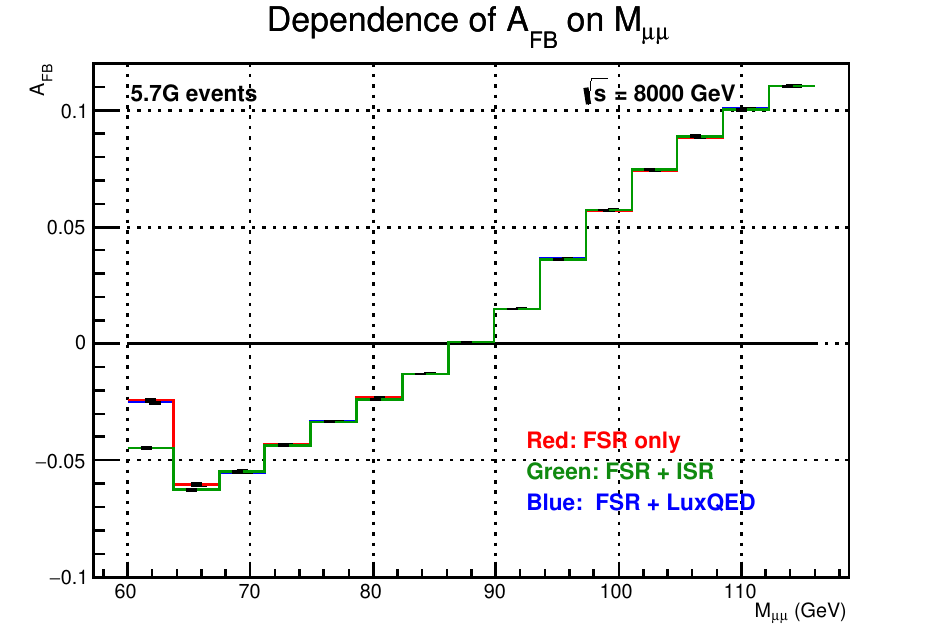}}
\put(3.0,0.5){\includegraphics[width=3.2in,height=2.6in]{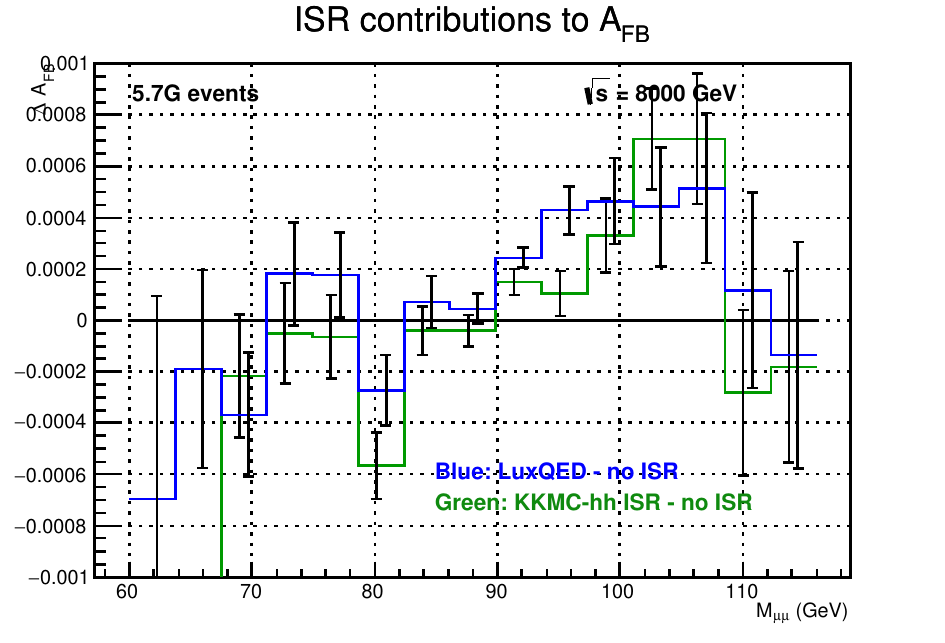}}
\end{picture}
\vspace{-0.75in}
\caption{Effect of ISR on \AFB\ in terms of dilepton mass.}
\end{center}
\end{figure}

\begin{figure}[H]
\begin{center}
\setlength{\unitlength}{1in}
\begin{picture}(6.5,3.0)
\put(0,0.5){\includegraphics[width=3.2in,height=2.6in]{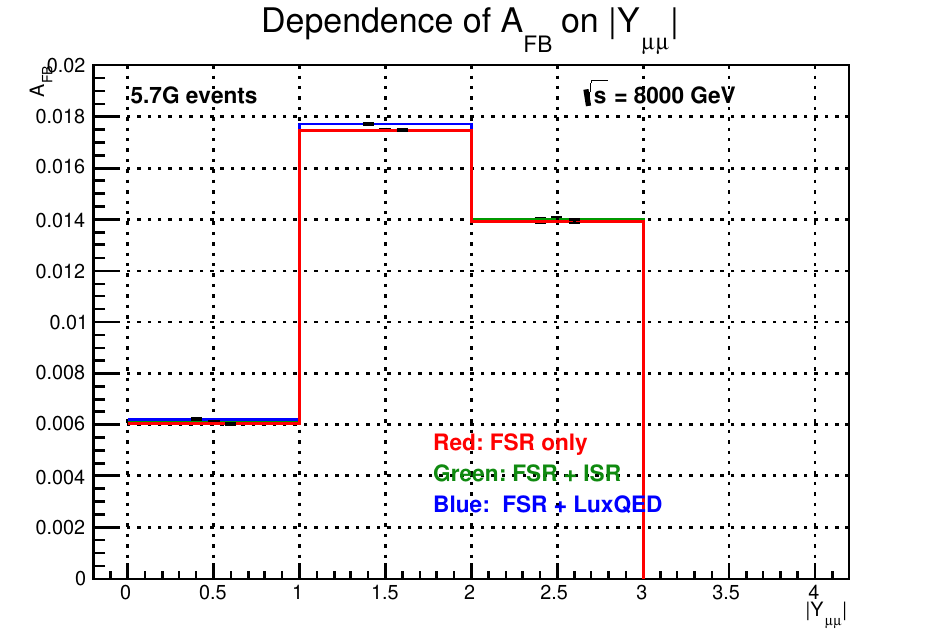}}
\put(3.0,0.5){\includegraphics[width=3.2in,height=2.6in]{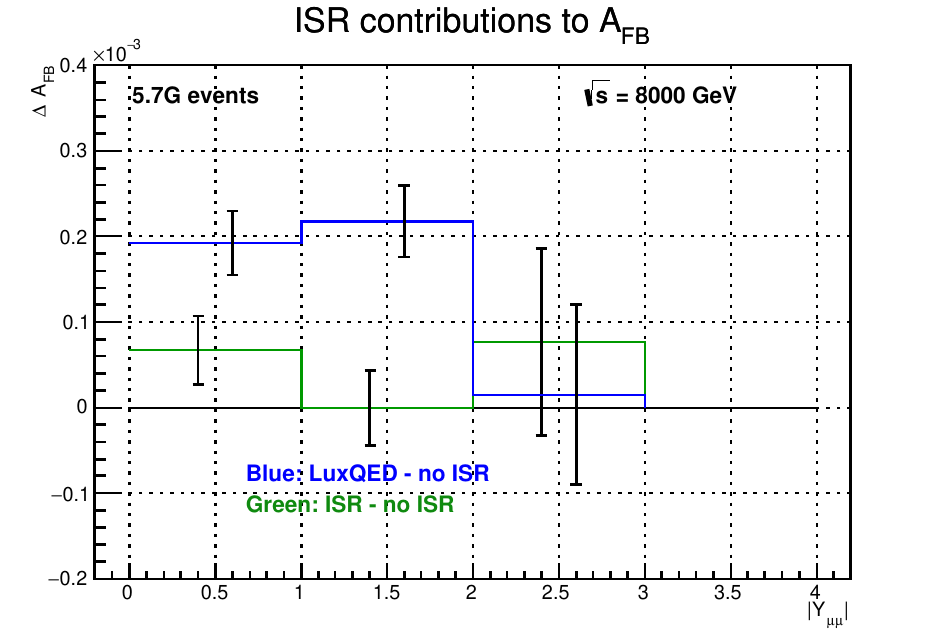}}
\end{picture}
\vspace{-0.75in}
\caption{Effect of ISR on \AFB\ in terms of dilepton rapidity.}
\end{center}
\end{figure}

Figures 5 and 6 show
the effect of ISR on $A_4$ as a function of the dilepton mass and rapidity.
In Fig. 6, the ISR contribution is again typically of order $10^{-3}$, and 
{\KK}MC-hh shows that it is approximately consistent with zero in the
vicinity of $M_Z$.  In Fig. 7, the ISR contribution from {\KK}MC-hh increases
for large rapidity, but is on the order of $10^{-4}$ for $Y_{ll} < 2$. The
LuxQED prediction is consistently below $5\times 10^{-4}$, but significantly
different from {\KK}MC-hh. 

\begin{figure}[ht!]
\begin{center}
\setlength{\unitlength}{1in}
\begin{picture}(6.5,3.0)
\put(0,0.5){\includegraphics[width=3.2in,height=2.6in]{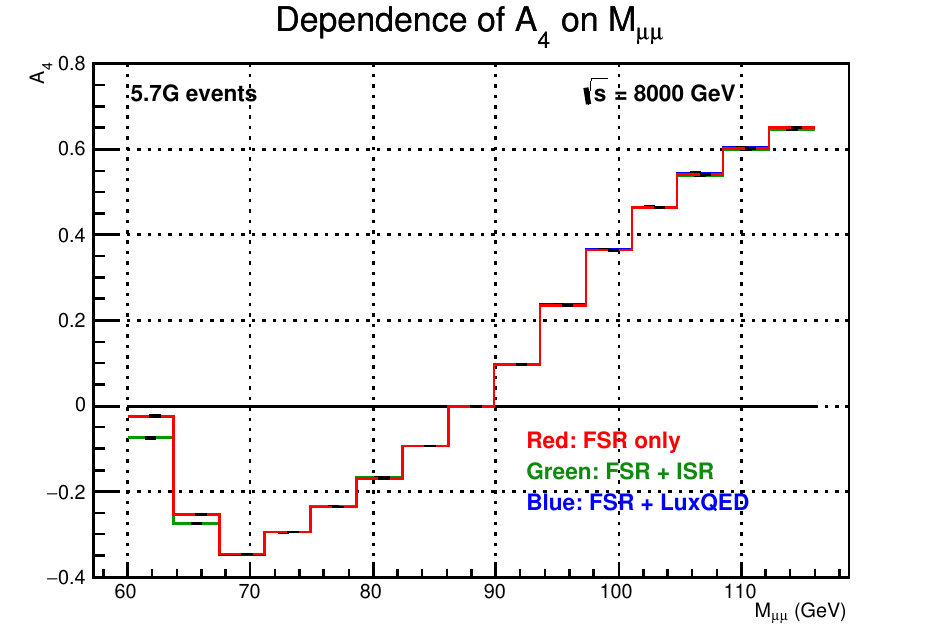}}
\put(3.0,0.5){\includegraphics[width=3.2in,height=2.6in]{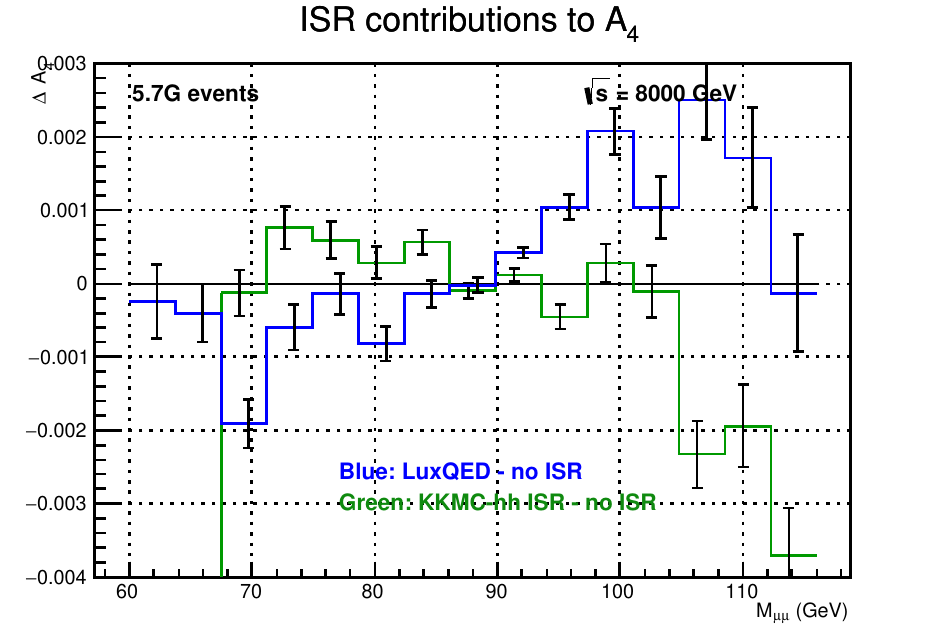}}
\end{picture}
\vspace{-0.75in}
\caption{Effect of ISR on $A_4$ in terms of dilepton mass.}
\end{center}
\end{figure}

\begin{figure}[ht!]
\begin{center}
\setlength{\unitlength}{1in}
\begin{picture}(6.5,3.0)
\put(0,0.5){\includegraphics[width=3.2in,height=2.6in]{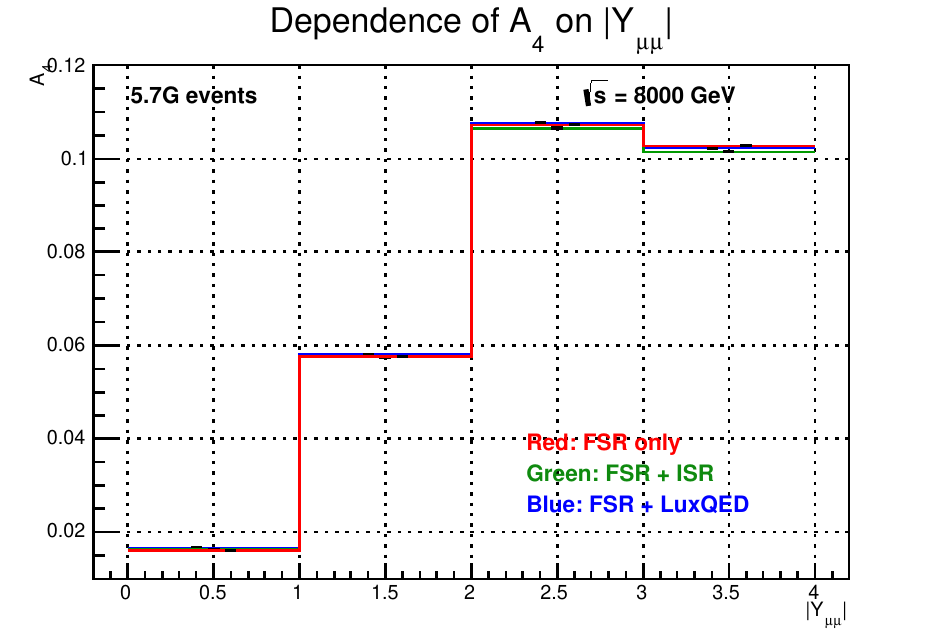}}
\put(3.0,0.5){\includegraphics[width=3.2in,height=2.6in]{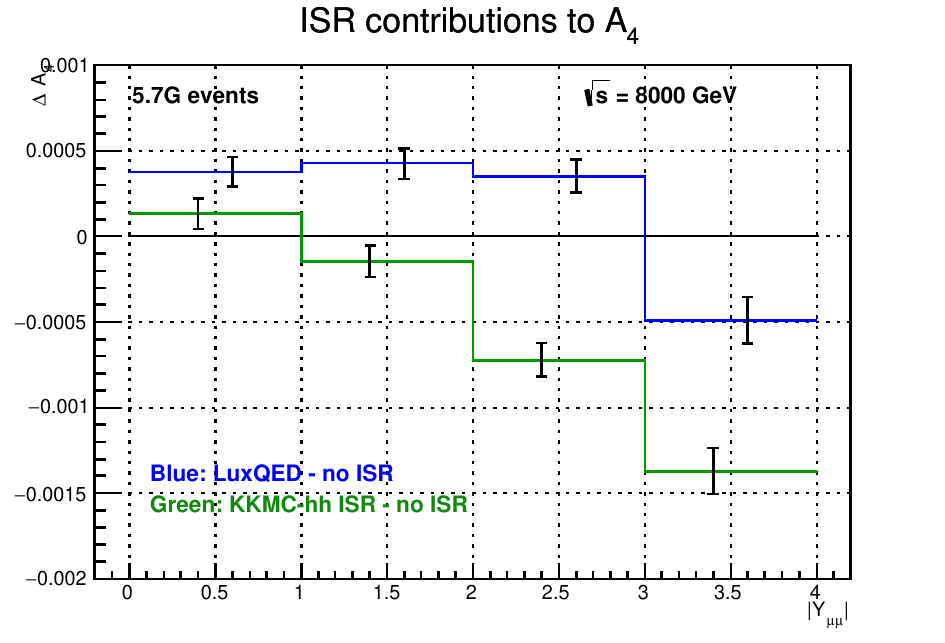}}
\end{picture}
\vspace{-0.75in}
\caption{Effect of ISR on $A_4$ in terms of dilepton rapidity.}
\end{center}
\end{figure}

\section{The Effect of Initial-Final Interference on Angular Distributions}

In this section, we consider the effect of quantum interference between 
initial and final state QED radiation (IFI) on the CS angular distributions,
forward-backward asymmetry, and $A_4$.  The use of \AFB\ or $A_4$ in 
determining the weak mixing angle is complicated by IFI, it 
a quantum uncertainty in any attempt to back out FSR from the measurement. 

All comparisons are without a QCD 
shower and use NNPDF3.1 parton distributions without QED corrections, since 
these are included in {\KK}MC-hh. The parameters are the same as in the 
previous section.

Table 2 shows the effect of ISR on the uncut and cut cross sections as well
as on the forward-backward asymmetry \AFB\ and on $A_4$. Both unshowered and 
showered results are available here, but the showered results are for a smaller
sample of $1.7\times 10^9$ events, so the precision is lower. The differences 
are shown relative to {\KK}MC-hh with both ISR and ISR on, but IFI off. For the
cross sections, percent differences are shown, while for the asymmetries, 
the differences are shown directly. The comparisons were calculated within a
single run by reweighting, and the errors take into account the weight
correlations, which reduce the uncertainty.

%\hspace{-1.5cm}
\vbox{
\begin{center}
\resizebox{\textwidth}{!}{%
\begin{tabular}{|l|c|c|c||c|c|c|}
\hline		& \multicolumn{3}{c||}{without shower}
		&\multicolumn{3}{c|}{with shower}\\
\hline
		& without IFI 	& with IFI 	& difference
		& without IFI 	& with IFI 	& difference\\
\hline
uncut $\sigma$ 	& 944.99(2)	& 944.91(2)     & $-0.0089(4)$\%
		& 938.46(4)	& 938.44(4)	& $-0.002(1)$\%\\
cut $\sigma$	& 442.36(1)	& 442.33(1)	& $-0.0070(5)$\%
		& 412.54(3)	& 412.56(3)	& $0.004(2)$\%\\
\AFB		& 0.01129(2)	& 0.01132(2)	&$(2.9\pm 1.1)\times 10^{-5}$
		& 0.01235(5)	& 0.02141(5)	&$(5.8\pm 2.6)\times 10^{-5}$\\
$A_4$		& 0.06057(3) 	& 0.06102(3)    &$(4.5\pm 0.3)\times 10^{-4}$
		& 0.06003(8)	& 0.06052(8)    &$(4.9\pm 0.8)\times 10^{-4}$\\
\hline
\end{tabular}
}
\\[1em]
{{\bf Table 2.} Effect of Initial-Final Interference, shown both with and 
without the HERWIG shower.}
\end{center}
}

The contribution of IFI on cross sections is very small, of the order 0.01\%,
while the effect of IFI on \AFB\ and $A_4$ is of order $10^{-5}$ and $10^{-4}$,
respectively. The presence of a QCD shower does not make a large difference in
the size of these effects. 

The IFI-dependence of the uncut and cut CS angle distributions
are shown in Fig. 7. The IFI contribution is shown
both with and without the QCD shower. The effect is typically a fraction of a 
per-mil, and angle-dependent. The uncut distribution (left) is relevant to 
$A_4$, and the cut distribution (right) is relevant to \AFB. 

\begin{figure}[hb!]
\begin{center}
\setlength{\unitlength}{1in}
\begin{picture}(6.5,3.0)
\put(0,0.5){\includegraphics[width=3.2in,height=2.6in]{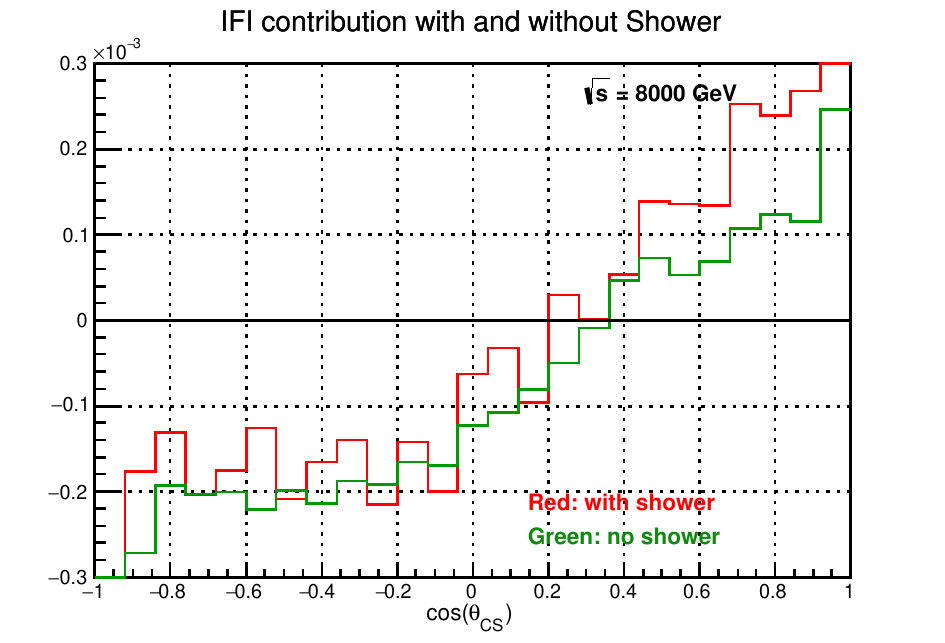}}
\put(3.0,0.5){\includegraphics[width=3.2in,height=2.6in]{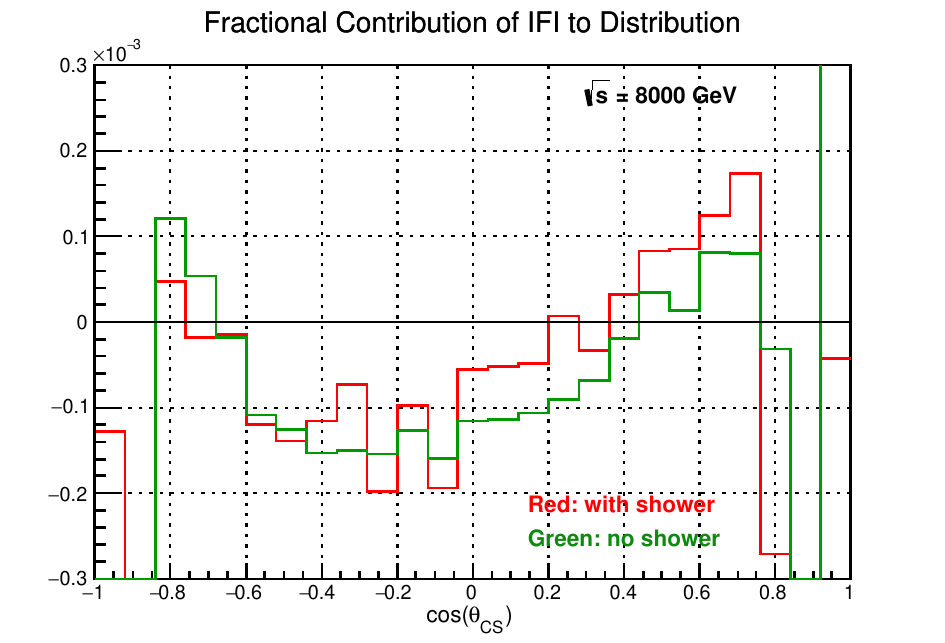}}
\end{picture}
\vspace{-0.75in}
\caption{Dependence of the Collins-Soper angular distribution on initial-final interference, without lepton cuts (left) and with them (right).}
\end{center}
\end{figure}

Fig. 8 shows the IFI effect on the forward-backward asymmetry, as a function of
the dilepton mass on the left, and the dilepton rapidity on the right. Fig. 9
shows similar comparisons for $A_4$. The IFI contribution to both \AFB\ and
$A_4$ is consistent with zero near $M_Z$ and at low dilepton rapidity. However,
$A_4$ is more sensitive to IFI than \AFB\ in general. The shower does not make
a large qualitative effect on the IFI contribution when binned in the dilepton
mass, but it enhances the IFI contribution somewhat when binned in rapidity.

\begin{figure}[ht!]
\begin{center}
\setlength{\unitlength}{1in}
\begin{picture}(6.5,3.0)
\put(0,0.5){\includegraphics[width=3.2in,height=2.6in]{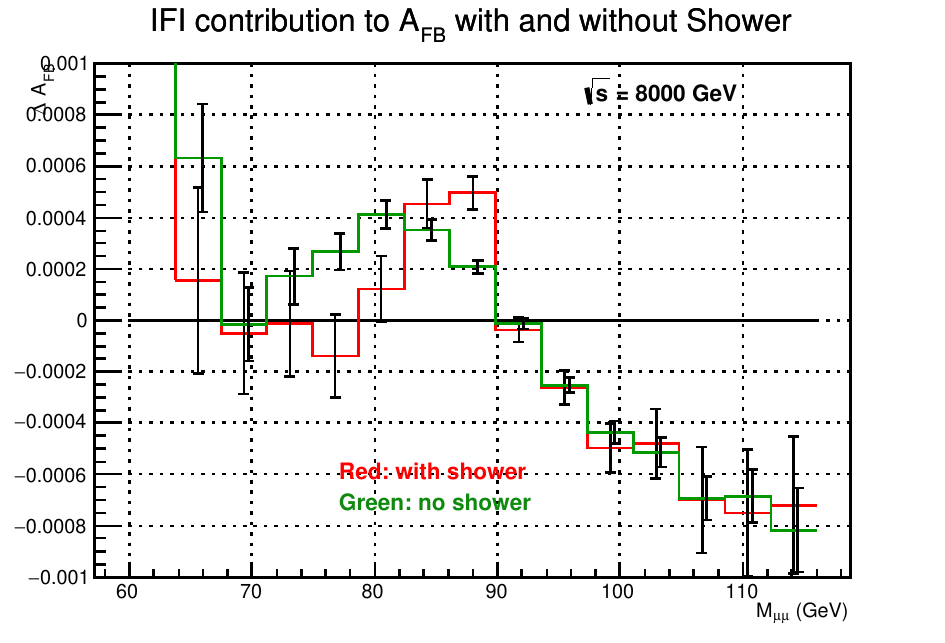}}
\put(3.0,0.5){\includegraphics[width=3.2in,height=2.6in]{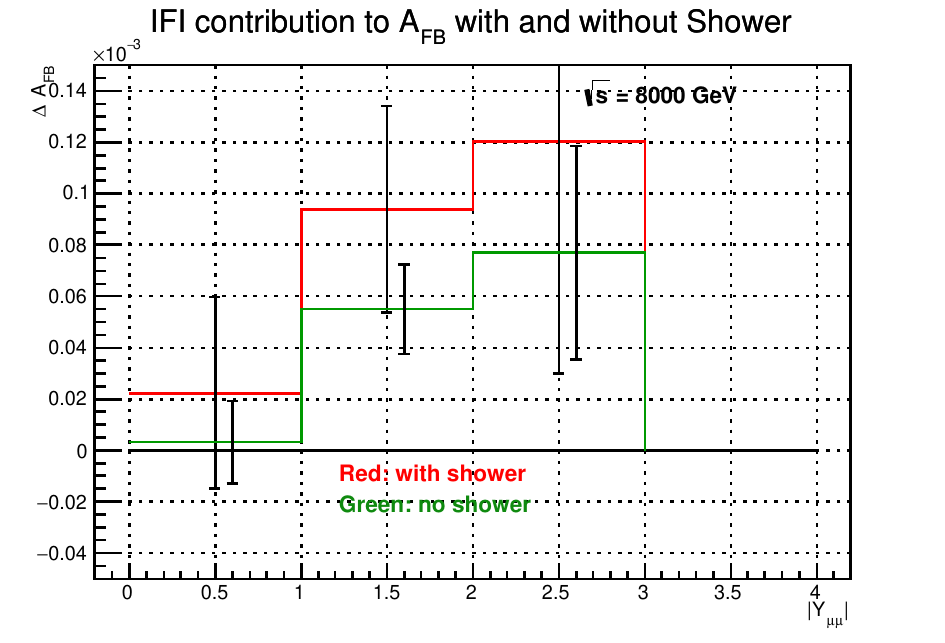}}
\end{picture}
\vspace{-0.75in}
\caption{The IFI contribution to \AFB\ as a function of $M_{ll}$  (left) and 
$Y_{ll}$ (right).}
\end{center}
\end{figure}

\begin{figure}[ht]
\begin{center}
\setlength{\unitlength}{1in}
\begin{picture}(6.5,3.0)
\put(0,0.5){\includegraphics[width=3.2in,height=2.6in]{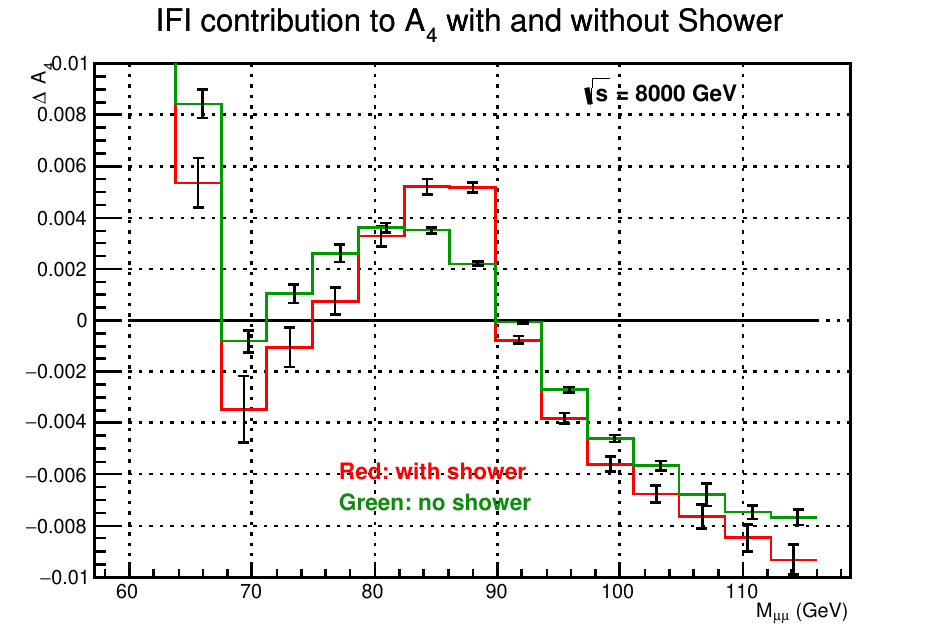}}
\put(3.0,0.5){\includegraphics[width=3.2in,height=2.6in]{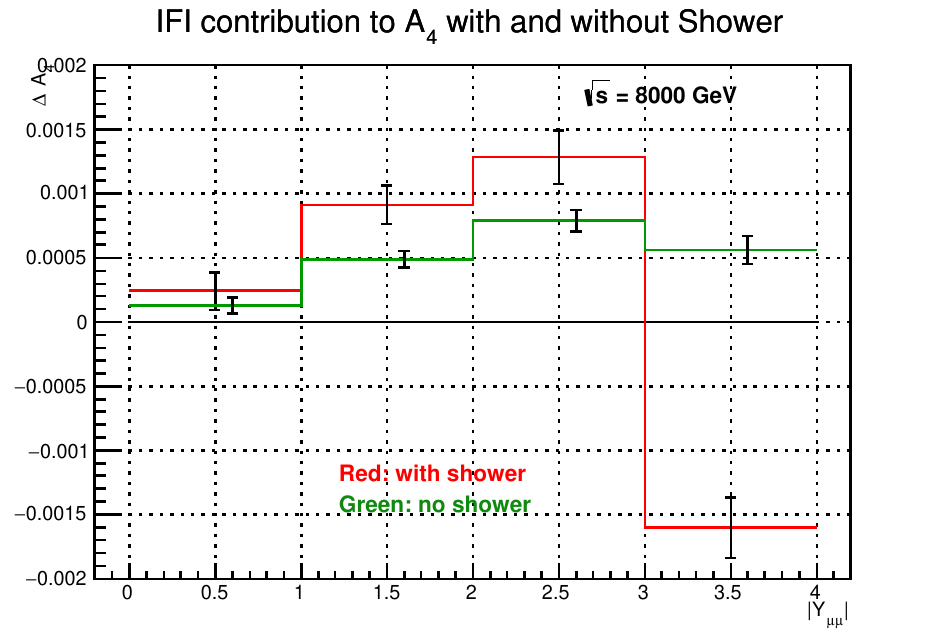}}
\end{picture}
\vspace{-0.75in}
\caption{The IFI contribution to $A_4$ as a function of $M_{ll}$ (left) and 
$Y_{ll}$ (right).}
\end{center}
\end{figure}

\section{Conclusions}

{\KK}MC-hh provides a precise tool for calculating exponentiated photonic
corrections to hadron scattering. We have presented estimates for the 
contributions of ISR and IFI to the \AFB\ and $A_4$ angular distributions
which will be useful for determining the weak mixing angle from LHC data. 
{\KK}MC-hh is particularly well suited to evaluating IFI due to its CEEX
exponentiation, which was developed in part to facilitate the calculation of 
interference effects. The ISR contribution is large enough that it cannot
be neglected in precision studies, and needs to be incorporated in some manner,
at least by including collinear photon emission in the PDFs, and preferably
by including exponentiated photon emission in the generator, as in {\KK}MC-hh. 

The {\em ab initio} calculation of QED emission from the quarks is unique 
to the approach of {\KK}MC-hh: other generators use calculations matched to
a QED-corrected PDF set. Studies comparing these approaches are in progress,
and the results will be interesting not just at the computational level, but
also conceptually, for better understanding the role of QED emission in
hadron scattering. 

Finally, we note that {\KK}MC-ee and {\KK}MC-hh are still under development. 
Thanks to the program's modular design, improvements in the pure
electroweak calculation can be readily incorporated in KKMC as they become
available.  Such an upgrade will be important in {\KK}MC-ee for future
$e^+e^-$ colliders\cite{Banerjee:2019mzj}, and {\KK}MC-hh will benefit at the
same time. In particular, an updated parametrization of $\alpha_{\rm QED,eff}$
\cite{Blondel:2019vdq} is available, as well as updated DIZET libraries
6.42\cite{Arbuzov:2005ma} and the recent version 6.45.

Tests of {\KK}MC-hh to date have focused on muon decays. {\KK}MC supports
$\tau$ lepton decays via TAUOLA\cite{tauola1993}, which has still needs to be 
tested in the context of {\KK}MC-hh to insure proper interplay with the shower. 
In addition, TAUOLA will eventually require an update, at least for future
$e^+e^-$ colliders, especially in the context of precision measurements of 
$\tau$ polarization effects.\cite{Banerjee:2019mle} Future results from 
Belle II~\cite{BelleII}  are likely to provide valuable input for reaching a
higher level of precision in modeling $\tau$ decays.

In the near future,
we expect to be able to address NLO QCD issues as well, at first by adding a
capability to add photonic corrections to events provided by any event 
generator, rather than running events generated by {\KK}MC-hh afterward. To
the extent that QCD and QED radiation factorize, which is true at leading log
and probably beyond that to some degree\cite{den-ditt1211.5078,dittmr}, the
two orders of showering should give equivalent results, but allowing the 
QCD shower to run run first increases the program's utility, 
and also provides a quantitative test of the factorization of QCD and QED 
radiation in this context.  Eventually, we 
anticipate incorporating NLO QCD internally, perhaps via the KrkNLO 
scheme.\cite{krknlo}

\end{document}